# Applications of the Variational Monte Carlo Method to the Ground States of the Lithium Atom and its Ions up to Z = 10 in the Presence of Magnetic Field

**S. B. Doma[1)], M. O. Shaker[2)], A. M. Farag[2)] and F. N. El-Gammal[3)]**

[1)]Mathematics Department, Faculty of Science, Alexandria University, Egypt
E-mail address: sbdoma@yahoo.com
[2)] Mathematics Department, Faculty of Science, Tanta University, Tanta, Egypt
[3)]Mathematics Department, Faculty of Science, Menofia University, Shebin El-Kom, Egypt

**Abstract**
The variational Monte Carlo method is applied to investigate the ground state energy of the lithium atom and its ions up to $Z = 10$ in the presence of an external magnetic field regime with $\gamma = 0 \sim 100$ a.u. Our calculations are based on using three forms of compact and accurate trial wave functions, which were put forward in calculating energies in the absence of magnetic field. The obtained results are in good agreement with the most recent accurate values and also with the exact values.

**Introduction**
Over the last decade continuing effort has gone into calculating, with ever increasing accuracy and with various methods, the energies of atoms and ions in neutron star magnetic fields. The motivation comes largely from the fact that features discovered [1–3] in the thermal emission spectra of isolated neutron stars may be due to absorption of photons by heavy atoms in the hot, thin atmospheres of these strongly magnetized cosmic objects [4]. Also, features of heavier elements may be present in the spectra of magnetic white dwarf stars [5, 6]. The comparison of the stationary transitions of the atom with the positions of the absorption edges of the observed spectrum yielded strong evidence for the existence of helium in the atmosphere of GD229 [7]. The most prominent of the unexplained magnetic objects is the white dwarf GD 229.

While comprehensive and precise data for hydrogen atom in strong magnetic fields have been available for some time (cf. [8–11]), this is less the case for atoms and ions with more than one electron. Recently significant progress has been achieved with respect to the interpretation of its rich spectrum. Extensive and precise calculations on the helium atom provided data for many excited states in a broad range of field strengths. The authors in [12] investigated the total electronic energies of the ground state and the first few excitations of the helium atom for the magnetic quantum number $M = 0$ and for both even and odd z-parity as well as their one-electron ionization energies as function of the magnetic field [12].

In early works the Coulomb field was considered in this limit actually as perturbation for a free electron in a super strong magnetic field. The motion of an electron parallel to the magnetic field is governed in the adiabatic approximation [13] by a 1D quasi-Coulomb potential with a parameter, dependent on the magnetic field strength. The detailed calculations of the hydrogen energy levels carried out by Rösner *et al.* [14] also retain the separation of the magnetic field strength domains due to decomposing the electronic wave function in terms of either spherical (for weak fields) or cylindrical harmonics (for strong fields). A powerful method to obtain comprehensive results on low-lying energy levels in the intermediate regime, in particular for the hydrogen atom, is provided by mesh methods [15]. Accurate calculations for arbitrary field strengths were carried out in refs. [16, 17] by the 2D mesh Hartree-Fock (HF) method. Investigations on the ground state as well as a number of excited states of helium, including the correlation energy, have recently been performed via a quantum Monte Carlo approach [18].

For atoms with several electrons the problem of the mixed symmetries is even more intricate than for hydrogen because different electrons feel very different Coulomb forces, i.e. possess different single-particle energies, and the domain of the intermediate fields, therefore, appears to be the sum of the intermediate domains for the separate electrons. Accurate ground state energies of atoms up to nuclear charge $Z = 10$ in the high-field regime were first determined by Ivanov and Schmelcher [19] who solved the 2D-HF equations on a flexible mesh.

There exist several investigations on the helium atom in the literature. Using a full configuration-interaction approach, which is based on a nonlinearly optimized anisotropic Gaussian basis set of one-particle functions, W. Becken and P. Schmelcher [20] calculated the total energies of the ground state and the first four excitations in each subspace as well as their one-electron ionization energies for the magnetic quantum number $M = -1$ and for both even and odd z-parity as well as singlet and triplet spin symmetry. Additionally they presented energies for electromagnetic transitions within the $M = -1$ subspace and between the $M = -1$ subspace and the $M = 0$ subspace. Also, W. Becken and P. Schmelcher investigated in ref. [21] the electromagnetic-transition probabilities for the helium atom embedded in a strong magnetic field in the complete regime $B = 0 \sim 100$ a.u. Furthermore, for the magnetic quantum numbers $M = 0, -1, -2, -3$ in the magnetic field regime B =100 ~10000 a.u., positive and negative z parity, singlet and triplet symmetry were carried out in refs. [22, 23]. A considerable number of higher angular-momentum states of the helium atom embedded in a magnetic field $B = 0 \sim 100$ a.u. was investigated. Spin singlet and triplet states with positive- and negative-z parity were considered for the magnetic quantum number $M = \pm 2$ and positive-z parity states were studied for $M = \pm 3$. Many of the excitations within these symmetries was also investigated in [24], where total energies, ionization energies as well as transition wave lengths were discussed in detail as function of the field strength. Employing a 2D mesh HF method Ivanov and Schmelcher [25-27] gave results of the ground and a few excited states of the carbon, beryllium and boron atom as well as $Be^+$ ion and $B^+$ ion in external uniform magnetic fields for field strengths ranging from zero up to $2.35 \times 10^9$ T with different spin projections $S_Z = -1, -2, -3$, $S_Z = 0, -1, -2$ and $S_Z = -1/2, -3/2, -5/2$, respectively. Also, the authors in [28] investigated the effects of a magnetic field with low to intermediate strength on several spectroscopic properties of the sodium atom, where ionisation energies, transition wave lengths and their dipole oscillator strengths were presented. Moreover, Ivanov and Schmelcher [29] presented precise HF results for several states in weak fields and quite satisfactory results for the intermediate region, where they applied a fully numerical 2D HF method to the problem of the Li atom in magnetic fields of arbitrary strength. This method enables one to perform calculations for various states and with approximately equal precision for weak, intermediate and super strong magnetic fields. Their main focus was the ground state of the Li atom and its ionization energies. To this end several electronic configurations of the Li atom and two configurations of the Li$^+$ ion are studied.

In ref. [30] Boblest *et al.* used the combination of a two-dimensional HF and a diffusion quantum Monte Carlo method in the presence of magnetic field strength to calculate ground state energy of lithium like ions up to $Z = 10$. Furthermore, the work of Jones *et al.* [31] for this atom is an interesting one, it contains calculations for the ground state and a few low-lying states of the Li atom at weak and intermediate fields.

In the present paper we have applied the variational Monte Carlo (VMC) method to study the three-electron system, by using three accurate trial wave functions. This method enables us performing calculations with approximately equal precision for weak and intermediate magnetic fields. Our main focus is the ground state of the lithium atom and its ions up to $Z = 10$. For this purpose, and extending to our previous works [32-34], in the present paper we have used the VMC method to compute the total energies and the corresponding standard deviations for the lithium atom with respect to the magnetic field strength. Our calculations were extended also to include the lithium ions up to $Z = 10$.

## 2. The Method of the Calculations

The VMC method is based on a combination of two ideas namely the variational principle and the Monte Carlo evaluation of integrals using importance sampling based on the Metropolis algorithm [35]. The VMC method is the simplest of the quantum Monte Carlo algorithms, and it is used to compute quantum expectation values of an operator with a given trial wave function $\Psi_T$.

In the VMC method, the ground state of a Hamiltonian $H$ is approximated by some trial wave function $|\Psi_T\rangle$. A number of parameters are introduced into $|\Psi_T\rangle$ and these parameters are varied to minimize the expectation value $E_{\Psi_T^2} = \langle \Psi_T|H|\Psi_T\rangle / \langle \Psi_T|\Psi_T\rangle$ in order to bring the trial wave function as close as possible to the actual ground state $|\Psi_0\rangle$. Wave functions of interacting systems are non-separable, and the integration needed to evaluate $E_{\Psi_T^2}$ is therefore a difficult task. Although it is possible to write these wave functions as linear combinations of separable terms, this tactic is viable only for a limited number of particles, since the length of such an expansion grows very quickly as the system size increases. The VMC method employs a stochastic integration that can treat the non-separable wave functions directly. The expectation value $E_{\Psi_T^2}$ [36] is written as

$$E_{\Psi_T^2} = \int \frac{|\Psi_T(R)|^2}{\langle \Psi_T|\Psi_T\rangle} \frac{[H\Psi_T](R)}{\Psi_T(R)} d^{3N}R \approx E_{VMC} = \frac{1}{N}\sum_{i=1}^{N} \frac{[H\Psi_T](R_i)}{\Psi_T(R_i)}, \qquad (2.1)$$

where $R = (R_1, R_2, R_3, \ldots, R_N)$ is a $3N$-dimensional vector encompassing the coordinates of all the $N$ particles in the system and the sum runs over $N$ such vectors $\{R_i\}$ sampled from the multivariate probability density $\rho(R) = \frac{|\Psi_T(R)|^2}{\langle \Psi_T|\Psi_T\rangle}$. The summand $E_L(R) = \frac{[H\Psi_T](R)}{\Psi_T(R)}$ is usually referred to as the local energy. We assume spin independent Hamiltonians, and therefore spin variables do not explicitly enter the evaluation of the expectation value (2.1). The trial wave function is supposed to be as close as possible to the true ground state wave function of the system, or more generally to an exact eigenstate of the Hamiltonian, if one is interested in studying also the properties of excited states. To calculate the integrals in Eq. (2.1) we firstly construct a trial wave function $\Psi_T^\alpha(R)$ depending on a set of $\alpha$-variational parameters $\alpha = (\alpha_1, \alpha_2, \ldots \ldots \ldots \alpha_N)$ and then vary the parameters to obtain the minimum energy.

In determining the expectation value of the local energy, $\langle E_L \rangle$, it is not necessary to carry out analytic integrations; and, since only differentiation of the trial wave function is required to evaluate the local energy, the trial wave function may take any desired functional form.

The variational energy $E_{VMC}$ is a stochastic variable, and an appropriate characterization of the random error $E_{VMC} - E_{\Psi_T^2}$ is thus an integral part of the VMC method. When the sampled local energies $E_L(\mathbf{R}_i)$ are sufficiently well behaved [37], this error can be represented by the variance of $E_{VMC}$. The weighted average $E_L(\mathbf{R})$ is evaluated at each point of the set of points $\{\mathbf{R}_i\}$. After a sufficient number of evaluations the VMC estimate of $E_{VMC}$ will be

$$E_{VMC} = \langle E_L \rangle_{\Psi_T^2} = \lim_{N \to \infty} \lim_{M \to \infty} \frac{1}{N} \frac{1}{M} \sum_{j=1}^{N} \sum_{i=1}^{M} E_L(\mathbf{R}_{ij}), \quad (2.2)$$

where, $M$ is the ensemble size of random numbers $\{\mathbf{R}_1, \mathbf{R}_1, \ldots \ldots, \mathbf{R}_M\}$, which may be generated by using a variety of methods [36] and $N$ is the number of ensembles. These ensembles so generated must reflect the distribution function itself. A given ensemble is chosen according to the Metropolis algorithm [35] and using random numbers. These random numbers may be generated by using a variety of methods [38]. When evaluating the energy of the system it is important to calculate the standard deviation [36]

$$\sigma = \sqrt{\frac{\langle E_L^2 \rangle_{\Psi_T^2} - \langle E_L \rangle^2_{\Psi_T^2}}{M(N-1)}} \quad (2.3)$$

of this energy. Since $\langle E_L \rangle_{\Psi_T^2}$ will be exact when an exact trial wave function is used, then the standard deviation of the local energy will be zero for this case [39]. Thus in the Monte Carlo method, the minimum of $\langle E_L \rangle_{\Psi_T^2}$ should coincide with a minimum in the standard deviation.

## 3. The Hamiltonian of the System

Our goal in this paper is to solve the Schrödinger equation for a lithium atom in an external magnetic field in order to calculate the ground-state energy eigenvalues as functions of the magnetic-field parameter. We first construct the Hamiltonian operator for the three-particle system in the absence of the external field. For this purpose we make use of the assumption of an infinitely heavy nucleus in the (unrestricted) HF approximation. The solution is established in Hylleraas coordinates [40] (with the $z$-axis oriented along the magnetic field). Hence, the non-relativistic Hamiltonian for the Lithium atom, in the absence of the field, under the Born-Oppenheimer approximation of zero order, that is, with the Li nucleus assumed to be of infinite mass, is (in Hartree-atomic units) [41]

$$H = -\sum_{i=1}^{3} \left( \frac{1}{2} \nabla_i^2 + \frac{Z}{r_i} \right) + \sum_{i=1}^{3} \sum_{j>i}^{3} \frac{1}{r_{ij}}, \quad (3.1)$$

where $\nabla_i$ is the 3-vector of the momentum of the $i$ th electron, $Z$ is the nuclear charge (here, $Z = 3$), $r_i$ is the distance between the $i$ th electron and the Li nucleus, and $r_{ij}$ are the interelectron distances.

In our calculations we used the form of $H$ in Hylleraas Coordinates [40] as follows:

$$H = -\frac{1}{2} \left( \sum_{i=1}^{3} \frac{\partial^2}{\partial r_i^2} + \sum_{i=1}^{3} \frac{2}{r_i} \frac{\partial}{\partial r_i} + \sum_{i<j}^{3} 2 \frac{\partial^2}{\partial r_{ij}^2} + \sum_{i<j}^{3} 2 \frac{\partial^2}{\partial r_{ij}^2} \sum_{i<j}^{3} \frac{4}{r_{ij}} \frac{\partial}{\partial r_{ij}} + \sum_{i \neq j}^{3} \frac{r_i + r_{ij} - r_j}{r_i r_{ij}} \frac{\partial^2}{\partial r_i \, \partial r_{ij}} + \sum_{i \neq j}^{3} \sum_{k>j}^{3} \frac{r_{ij} + r_{ik} + r_{jk}}{r_{ij} r_{ik}} \frac{\partial^2}{\partial r_{ij} \, \partial r_{ik}} \right) + \sum_{i=1}^{3} \frac{-Z}{r_i} + \sum_{1=i<j}^{3} \frac{1}{r_{ij}}, \quad (3.2)$$

According to the assumption that the nuclear mass is infinite and the magnetic field is oriented along the $z$-axis, the non-relativistic Hamiltonian $\mathcal{H}$ for the Li atom in the presence of the magnetic field (in atomic units (a. u.)) takes the form

$$\mathcal{H} = H + \left[\frac{\gamma^2 \rho^2}{8} + \frac{\gamma(L_z + 2S_z)}{2}\right]. \quad (3.3)$$

where $\rho^2 = (x_1^2 + y_1^2) + (x_2^2 + y_2^2) + (x_3^2 + y_3^2)$, $\gamma$ is the magnetic field parameter, $S_z$ is the $z$-component of the total spin, $L_z$ is the $z$-component of the total angular momentum, $\frac{\gamma^2 \rho^2}{8}$ is the diamagnetic term, $\frac{\gamma}{2}L_z$ is the Zeeman term, $-\frac{Z}{r_1} - \frac{Z}{r_2} - \frac{Z}{r_3}$ are the attractive Coulomb interactions with the nucleus and $\gamma S_z$ represents the spin Zeeman term.

**4. The Trial Wave Functions**

In the VMC method, the accuracy of the trial wave function determines directly the accuracy in the energy obtained in the calculation. The choice of trial wave function $\Psi_T(R)$ is critical in VMC calculation. How to choose it is however a highly non-trivial task. The trial wave function must approximate an exact eigenstate in order that accurate results are to be obtained. Also, the trial wave function improves the importance sampling and reduces the cost of obtaining a certain statistical accuracy. A good trial wave function should exhibit much of the same features as does the exact wave function. One possible guideline in choosing the trial wave function is the use of the constraints about the behavior of the wave function when the distance between one electron and the nucleus or two electron approaches zero. These constraints are called "cusp conditions" and are related to the derivative of the wave function. More details about the trial wave function can be found in [42].

Usually the correlated wave function, $\psi$, used in the VMC method is built in the form of product of a symmetric correlation factor, $f$, which includes the dynamic correlation among the electrons, times a model wave function, that provides the correct properties of the exact wave function such as the spin and the angular momentum of the atom, and is antisymmetric in the electronic coordinates. Physical relevance arguments are followed to choose the trial wave function (see, e.g., Turbiner [43]). In particular, we construct wave functions that allow us to reproduce both the Coulomb singularities in $r_i$ and in $r_{ij}$ and the correct asymptotic behavior of large distances. As a result, the wave function of the $2_{S_{1/2}}$ Li ground state is written in the particular form

$$\psi = A[\emptyset(r_1, r_2, r_3)], \quad (4.1)$$

with trial functions, which were examined, consisting of exponentials in all the relative coordinates, in some cases, multiplied by pre-exponential factors dependent linearly on the interparticle distances, in the form

$$\phi(r_1, r_2, r_3) = f(r_1, r_2, r_3, r_{12}, r_{13}, r_{23}) e^{-\alpha_1 r_1 - \alpha_2 r_2 - \alpha_3 r_3 - \alpha_{12} r_{12} - \alpha_{13} r_{13} - \alpha_{23} r_{23}}. \quad (4.2)$$

In Eq. (4.2) the pre-exponential factor is a linear function of its arguments, whereas $\alpha_i$ and $\alpha_{ij}$ are nonlinear parameters. $A$ is the three-particle antisymmetrizer

$$A = I - P_{12} - P_{13} - P_{23} + P_{231} + P_{312}. \quad (4.3)$$

Here, $P_{ij}$ represents the permutation $i \leftrightarrow j$, and $P_{ijk}$ stands for the permutation of 123 into $ijk$. In total, the function $\psi$ of Eq. (4.1) is characterized by six parameters, plus any parameters that may occur in the pre-exponential factor $f$.

In our calculations we have used two accurate wave functions $\Phi_1$ and $\Phi_2$ [44], for solving the Schrödinger equation to obtain the energies of the lithium atom, which take the form:

$$\Phi_1 = (1 + \beta_1 r_3 + \gamma_1 r_{13})e^{-\alpha_1 r_1 - \alpha_2 r_2 - \alpha_3 r_3 - \alpha_{12} r_{12} - \alpha_{13} r_{13} - \alpha_{23} r_{23}} \quad (4.4)$$

$$\Phi_2 = e^{-\alpha_1 r_1 - \alpha_2 r_2 - \alpha_3 r_3 - \alpha_{12} r_{12} - \alpha_{13} r_{13} - \alpha_{23} r_{23}} \quad (4.5)$$

Moreover, we have obtained the energies of the lithium ions up to $Z = 10$ by using an accurate wave function $\phi_3$ which were put forward in calculating energies in the absence of the magnetic field [44]. This function takes the form:

$$\Phi_3 = (1 + \beta_1 r_3)e^{-\alpha_1 r_1 - \alpha_2 r_2 - \alpha_3 r_3 - \alpha_{12} r_{12} - \alpha_{13} r_{13} - \alpha_{23} r_{23}} \quad (4.6)$$

In our calculations we took the values of the variational parameters in $\Phi_1, \Phi_2$ and $\Phi_3$ from Ref [44].

**5. Results**
The Monte Carlo method described here has been employed for calculating the ground state energies of the lithium atom in the magnetic field regime between 0 a.u. and 100 a.u. as well as the lithium like ions up to $Z = 10$. All energies are obtained in atomic units ($m_e = e = \hbar = 1$), i.e. in Hartrees (Ha), such that (1 Ha = 1 atomic unit), with set of $1 \times 10^7$ Monte Carlo integration points in order to make the statistical error as low as possible. With increasing field strength this state undergoes two transitions involving three different electronic configurations. For weak fields up to $\gamma = 0.17633$ the ground-state arises from the field-free $1s^2 2s$ configuration. For intermediate fields ($0.17633 < \gamma < 2.071814$) the ground state is constituted by the $1s^2 2p_{-1}$ configuration and for $\gamma > 2.071814$ the ground state configuration is $1s2p_{-1}3d_{-2}$. The change of the ground state configuration takes place at $\gamma = 2.071814$. It is clear that the state $1s^2 2s$ of the lithium atom is the ground state only for relatively weak fields.

For each separate configuration, the effect of the increasing field strength consists in compressing the electronic distribution towards the z axis. For the $1s2p_{-1}3d_{-2}$ configuration, for which all single electron binding energies increase unlimited for $\gamma \to \infty$, a shrinking process of this distribution in the z direction is also visible. For the $1s^2 2p_{-1}$ configuration this effect is not distinct for the relevant field strengths. For the $1s^2 2s$ state the opposite effect can be observed: the 2s electronic charge distribution along the z-axis expands slightly in weak magnetic fields.

For the ground state, the Hamiltonian integrals of Li are easily done when the wave function is given by Eq. (4.4). Guevara *et al.* [44] used this trial wave function to calculate energies for the ground-state of the lithium atom in the absence of magnetic field. They could obtained very accurate results compared to the corresponding exact values.

In the presence of a magnetic field an atom and its physics is subject to a variety of changes. For instance, the conserved quantum numbers are reduced to the total angular momentum z-projection M, the total z-parity $\Pi_z$, the total spin z-projection $S_z$ and the total spin $S^2$. Since the wave functions and energies of the states strongly depend on the magnetic field strength, the ground state energy and configuration are also affected by the magnetic field.

## 5.1 The ground state of the lithium atom

The only work on the Li atom in a magnetic field with which we can compare our results is Ref. [31]. In this reference a HF calculations were performed for weak and intermediate magnetic field strengths. For the lithium atom ($Z = 3$), we have calculated the total energies of the ground state as functions of the magnetic field. The energy of the ground state of the lithium atom in the absence of magnetic field is -7.432748 (Ha), which is closer to -7.43275 (Ha) obtained by Jones et al. [31].

The Monte Carlo process described here has been employed for the ground state of the lithium atom. Our calculations for the ground-state of lithium atom are based on using two accurate trial wave functions $\Phi_1$ and $\Phi_2$. In Table-1 we present our results for the behavior of the total energy at different values for the magnetic field strength $\gamma$. We end our results in Table-1 with a comparison with the findings of Ivanov and Jones [29, 31]. Our energy values coincide with those of ref. [31] for weak fields $0 < \gamma < 0.17633$ and lie substantially low in the intermediate regime.

It is seen from Table-1 that the energy of the lithium atom in the absence of magnetic field is -7.432748 (Ha), which is closer to the value -7.4327 (Ha) obtained by Jones et al. [31]. Since Ivanov et al.'s results [29] are rounded, one cannot come to a conclusion that their values for field strengths $\gamma = 1.8, 3.0, 5.4, 10, 50$ and $\gamma = 100$ are lower than our results with respect to the trial wave function $\Phi_1$.

Figure-1 shows the variation of the ground state total energy with respect to the magnetic field strength $\gamma$ from $\gamma = 0$ to $\gamma = 100$ using the trial wave function $\Phi_1$. It is obvious that the total energy increases monotonically with increasing the magnetic field strength because we can expect from the quantum mechanical perturbation theory that the energy of the atom increased with increasing $\gamma$ when $\gamma$ is small, the behavior of E in the present method is more reasonable. Furthermore, Figure-2 shows the variation of the ground state total energy with respect to the magnetic field strength $\gamma$ from $\gamma = 0$ to $\gamma = 100$ by using the second trial wave function $\Phi_2$.

As can be seen from Table-1, the ground state total energy is raised from -7.432748 (Ha) at $\gamma = 0$ to 71.83664(Ha) at $\gamma = 100$ a.u. with respect to the trial wave function $\Phi_1$. Furthermore, the total energy is raised from -7.37567 (Ha) at $\gamma = 0$ to 71.75645 (Ha) at $\gamma = 100$ a.u. with respect to the trial wave function $\Phi_2$. This state is the most tightly bound state for all field strengths because the electrons in this state are much closer to the nucleus than in other states.

Table-1 Total energies of the ground state of the Li atom, in Ha, obtained by using the two wave functions of Eq. (4.4) and Eq. (4.5) in the regime of field strength $\gamma = 0, \ldots, 100$. The standard deviations, $\sigma$, of our results are also given.

| $\gamma$ | Present work | | Other works | |
|---|---|---|---|---|
| | $E(\Phi_1)$ $\sigma$ | $E(\Phi_2)$ $\sigma$ | $E^{[29]}$ | $E^{[31]}$ |
| 0.0000 | -7.432748 $7 \times 10^{-5}$ | -7.37567 $9 \times 10^{-5}$ | -7.43275 | -7.4327 |
| 0.0010 | -7.433196 $5 \times 10^{-5}$ | -7.37596 $5 \times 10^{-5}$ | -7.43326 | |
| 0.0018 | -7.433624 $4 \times 10^{-5}$ | -7.37624 $4 \times 10^{-5}$ | -7.43365 | -7.4337 |
| 0.0020 | -7.433893 $2 \times 10^{-5}$ | -7.37753 $7 \times 10^{-5}$ | -7.43375 | |
| 0.0050 | -7.435158 | -7.37784 | -7.43522 | |
|          | $5 \times 10^{-4}$       | $5 \times 10^{-5}$       |          |         |
| 0.0090   | -7.437062 $3 \times 10^{-4}$ | -7.37876 $4 \times 10^{-5}$ | -7.43713 | -7.4371 |
| 0.0100   | -7.437693 $7 \times 10^{-4}$ | -7.37913 $2 \times 10^{-5}$ | -7.43760 |         |
| 0.0180   | -7.441323 $5 \times 10^{-4}$ | -7.38143 $5 \times 10^{-4}$ | -7.44125 | -7.4412 |
| 0.0200   | -7.442008 $4 \times 10^{-4}$ | -7.38208 $3 \times 10^{-4}$ | -7.44214 |         |
| 0.0500   | -7.453947 $7 \times 10^{-4}$ | -7.38347 $5 \times 10^{-4}$ | -7.45398 |         |
| 0.0540   | -7.455361 $5 \times 10^{-4}$ | -7.395361 $5 \times 10^{-4}$ | -7.45537 | -7.4553 |
| 0.1000   | -7.468596 $6 \times 10^{-4}$ | -7.39578 $4 \times 10^{-4}$ | -7.46857 |         |
| 0.1260   | -7.473900 $5 \times 10^{-4}$ | -7.396743 $7 \times 10^{-4}$ | -7.47408 | -7.4739 |
| 0.17633  | -7.481644 $3 \times 10^{-4}$ | -7.397582 $5 \times 10^{-4}$ | -7.48162 |         |
| 0.1800   | -7.482021 $5 \times 10^{-4}$ | -7.398864 $6 \times 10^{-4}$ | -7.48204 | -7.4814 |
| 0.2000   | -7.484008 $3 \times 10^{-4}$ | -7.399250 $5 \times 10^{-4}$ | -7.48400 |         |
| 0.5000   | -7.47740 $5 \times 10^{-4}$  | -7.398655 $5 \times 10^{-4}$ | -7.47741 |         |
| 0.5400   | -7.473516 $8 \times 10^{-4}$ | -7.398572 $7 \times 10^{-4}$ | -7.47351 | -7.4731 |
| 0.9000   | -7.421701 $5 \times 10^{-4}$ | -7.386485 $6 \times 10^{-4}$ | -7.42504 | -7.4240 |
| 1.0000   | -7.406423 $3 \times 10^{-4}$ | -7.387954 $3 \times 10^{-4}$ | -7.40879 |         |
| 1.2600   | -7.358321 $4 \times 10^{-4}$ | -7.278541 $6 \times 10^{-4}$ | -7.36226 | -7.3609 |
| 1.8000   | -7.248103 $3 \times 10^{-4}$ | -7.178943 $5 \times 10^{-4}$ | -7.24603 | -7.2446 |
| 2.0000   | -7.195834 $2 \times 10^{-4}$ | -7.09543 $3 \times 10^{-4}$  | -7.19621 |         |
| 2.071814 | -7.176851 $3 \times 10^{-4}$ | -7.076534 $4 \times 10^{-4}$ | -7.17745 |         |
| 2.5000   | -7.047236 $3 \times 10^{-4}$ | -7.023864 $4 \times 10^{-4}$ | -7.05619 |         |
| 3.0000   | -6.896353 $4 \times 10^{-4}$ | -6.546783 $5 \times 10^{-4}$ | -6.89559 |         |
| 3.6000   | -6.676372 $5 \times 10^{-4}$ | -6.579452 $5 \times 10^{-4}$ | -6.67874 | -6.6640 |
| 5.0000   | -6.087652 $1 \times 10^{-4}$ | -6.016542 $3 \times 10^{-4}$ | -6.08811 |         |
| 5.4000   | -5.902153 $6 \times 10^{-4}$ | -5.863524 $7 \times 10^{-4}$ | -5.90113 | -5.8772 |
| 7.0000   | -5.087145                | -5.077438                | -5.08909 |         |
|     | $2 \times 10^{-4}$    | $4 \times 10^{-4}$   |          |   |
| 10. | -3.358745             | -3.13743             | -3.35777 |   |
|     | $4 \times 10^{-4}$    | $6 \times 10^{-4}$   |          |   |
| 20. | 3.406354              | 3.36473              | 3.49120  |   |
|     | $6 \times 10^{-4}$    | $7 \times 10^{-4}$   |          |   |
| 50. | 27.78853              | 27.55477             | 27.6916  |   |
|     | $6 \times 10^{-4}$    | $9 \times 10^{-4}$   |          |   |
| 100.| 71.83664              | 71.75645             | 71.807   |   |
|     | $5 \times 10^{-4}$    | $6 \times 10^{-4}$   |          |   |

[29]: Total energies obtained from Ref [29]; [31]: total energies obtained from Ref [31].

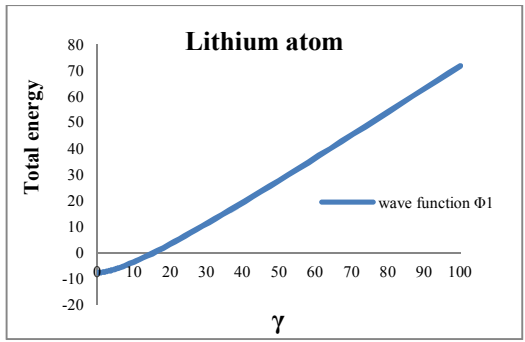
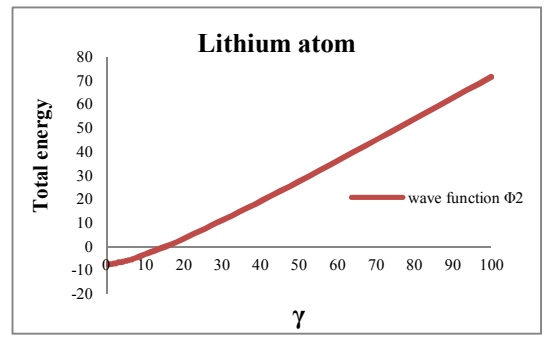

Figure-1 The ground state-total energy versus the magnetic field strength $\gamma$/a.u. from $\gamma = 0$ to $\gamma = 100$ using the trial wave function $\Phi_1$.

Figure-2 The ground state total energy versus the magnetic field strength $\gamma$/a.u. from $\gamma = 0$ to $\gamma = 100$ using the trial wave function $\Phi_2$.

## 5.2 The ground state of the lithium like ions up to $Z = 10$

We have calculated the total energies of the lithium like ions up to $Z = 10$ as functions of the magnetic field by using the trial wave function $\phi_3$. In Table-2 we present the total energy of the lithium like ions. In most cases of Table-2, the obtained results are in good agreement with the exact values. Also, the associated standard deviations have very small values, which vary between $10^{-5}$ and $10^{-4}$, this is due to the large number of MC points. It is clear from Table-2 that our result for the lithium atom ($Z = 3$) is -7.69085, which is more accurate than the others.

Figure-3 shows the behavior of the total energies obtained for the variational wave function of the form given in Eq. (4.6) versus lithium like ions up to $Z = 10$. All energies are obtained in Ha. We end these presentations with a comparison between our results and the results of Boblest and Schimeczek [30] in Table-2.

Table-2 Magnetic field strengths $\gamma$ and energy values E (in Ha) with the ground state-configuration change for lithium-like ions. The standard deviations, $\sigma$, of our results are also given.

| Z | $\gamma$ | E (Ha) $\sigma$ Our work | E[30](Ha) |
|---|---|---|---|
| 3 | 0.12212 | -7.69085  $1 \times 10^{-4}$ | -7.6905 |
| 4 | 0.14229 | -15.0576  $4 \times 10^{-5}$ | -15.0670 |
| 5 | 0.15510 | -24.9326  $4 \times 10^{-5}$ | -24.9386 |
| 6 | 0.16404 | -37.3084  $3 \times 10^{-5}$ | -37.309 |
| 7 | 0.17059 | -52.1764  $6 \times 10^{-5}$ | -52.184 |
| 8 | 0.17556 | -69.545  $4 \times 10^{-5}$ | -69.551 |
| 9 | 0.17947 | -89.114  $7 \times 10^{-6}$ | -89.418 |
| 10 | 0.18266 | -111.653  $6 \times 10^{-6}$ | -111.783 |

[30]: Total energies obtained from Ref [30].

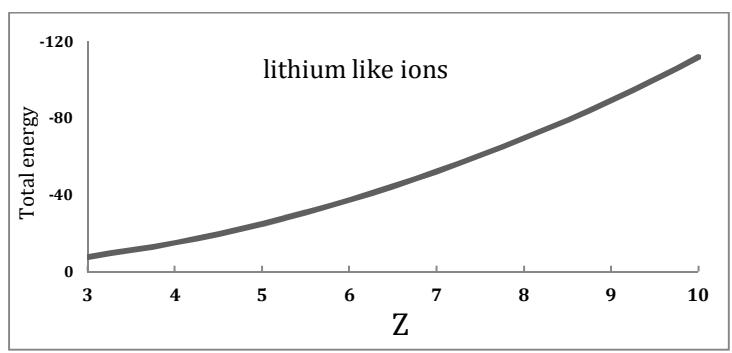

Figure-3 Total energies of the Lithium like ions versus Z, up to $Z = 10$.

## 6. Conclusions

We have applied the VMC method to a magnetized Li atom in order to calculate the ground state energy in the magnetic field regime $\gamma = 0 \sim 100$ a.u. by using simple and compact few-parameter trial wave functions. Moreover, we calculated the total energies for the lithium likeions up to $Z = 10$ in a magnetic field. The used functions $\Phi_1, \Phi_2$ and $\Phi_3$ are the most accurate among several existing few-parameter trial wave functions for the lithium atom and its ions up to $Z = 10$, respectively. Moreover, our results for the total energies with respect to the magnetic field provide high accuracy results of the ground state energy in the magnetic field regime $\gamma = 0 \sim 100$ a.u. and are in good agreement with the most recent accurate values. The energies were plotted as function of the magnetic field strengths $\gamma$ to show graphically the effect of $\gamma$ on the total energy.

Our results on the total energies are illustrated in figures 1, 2 and 3. These figures show in particular the ground state energies for the different regimes of the field strength. Finally, we conclude that the presented analysis represents a starting step in the mathematical treatment of this quantum-mechanical problem, since for example we are in need to examine other forms for the pre-factor term $f$ which take into considerations the influence of all the electron relative coordinates in the trial wave function. Also, we need to examine other forms of the trial wave functions in the application of the variational Monte Carlo method in such problems.